\documentstyle[12pt]{article}

\large
\newcommand{\be}{\begin{eqnarray}}
\newcommand{\ee}{\end{eqnarray}}

\begin{document}
\setlength{\baselineskip}{21pt}
\pagestyle{empty}
\vfill
\eject
\begin{flushright}
SUNY-NTG-94/6
\end{flushright}

\vskip 2.0cm
\centerline{\bf Spectral sum rules and Selberg's integral formula}
\vskip 2.0 cm
\centerline{Jacobus Verbaarschot}
\vskip .2cm
\centerline{Department of Physics}
\centerline{SUNY, Stony Brook, New York 11794}
\vskip 2cm

\centerline{\bf Abstract}
Using Selberg's integral formula we derive all Leutwyler-Smilga type
sum rules for one and two flavors, and for each of the three chiral
random matrix ensembles. In agreement with arguments from effective
field theory, all sum rules for $N_f = 1$ coincide for the three
ensembles. The connection between spectral correlations
and the low-energy effective partition function is discussed.

\vfill
\noindent
\begin{flushleft}
SUNY-NTG-94/6\\
January 1994
\end{flushleft}
\eject
\pagestyle{plain}

\vskip 1.5cm
\noindent{\bf 1. Introduction}
\setcounter{equation}{0}
\vskip 0.5 cm
As has been shown by Leutwyler and Smilga \cite{LEUTWYLER-SMILGA-1993},
chiral perturbation theory of the low-energy effective QCD partition function
leads to sum rules for the inverse powers of the
eigenvalues of the Dirac operator in QCD. On the other hand these sum rules
can also be derived from chiral random matrix theory
\cite{SHURYAK-VERBAARSCHOT-1993,VERBAARSCHOT-1994P} with only
the $symmetries$ of the QCD Dirac operator as input. They follow from the
microscopic correlations between the eigenvalues near zero virtuality, which
are conjectured to be universal
\cite{SHURYAK-VERBAARSCHOT-1993,VERBAARSCHOT-ZAHED-1993}.
In \cite{VERBAARSCHOT-1994P} we have argued
that there are three different universality classes: $SU(2)$-gauge theory
with fundamental fermions, $SU(N_c)$, $N_c \ge 3$
with fundamental fermions and $SU(N_c)$, $N_c \ge 2$
with adjoint fermions. The Dirac operator can be chosen  real,
is complex or can be regrouped into quaternions, respectively. In analogy
with the classical random matrix ensembles \cite{DYSON-1962}, the corresponding
chiral random matrix
theory will be called the chiral orthogonal ensemble (chGOE), the chiral
unitary ensemble (chGUE) and the chiral symplectic ensemble (chGSE).
As was shown in
\cite{VERBAARSCHOT-1994P} the spectral correlations of
chiral random matrix theory are different in each of the three cases.
Indeed, for two or more flavors the effective theory is also different
\cite{LEUTWYLER-SMILGA-1993,DIAKONOV-PETROV-1993}.
However, for $one$ flavor,
the low-energy effective partition function, with proper identification
of the parameters,  is
the same in each of the three cases \cite{LEUTWYLER-SMILGA-1993}, and therefore
also the spectral sum rules are the same.
This leads to the paradox \cite{SMILGA-1994}
of how an infinite family of sum rules can be the
same, whereas the spectrum near zero virtuality is different.
This issue will be resolved below by an explicit calculation of all sum rules.

In general,
spectral correlation functions can be calculated exactly, but the
integrals for the sum-rules can be evaluated
analytically only for the simplest correlation functions
\cite{VERBAARSCHOT-ZAHED-1993}.
However, there is a more direct way to obtain the same results. The spectral
sum rules can be expressed in terms of the joint eigenvalue density. The
resulting integrals are very complicated, but fortunately, exactly
these types of integrals have been considered before by Selberg
\cite{SELBERG-1944} and can
easily be generalized with the help of Aomoto's proof \cite{AOMOTO-1988}
thereof (see \cite{MEHTA-1991} for a detailed discussion).
Using this technique we are able to derive $all$ sum rules for one and two
flavors.

\vskip 1.5cm
\noindent{\bf 2. Leutwyler-Smilga spectral sum rules}
\vskip 0.5 cm
For $N_f$
flavors with masses $m_f$ ($m_f \rightarrow 0$) the QCD partition function
in the sector with $\nu$ zero modes is defined by
\be
Z_\nu^{\rm QCD}(m) = \langle \prod^{N_f}_{f=1}\prod_{\lambda_n>0}
(\lambda_n^2 + m_f^2) m_f^{\nu}\rangle_{S_\nu(A)},
\ee
where the average is over gauge field configurations
with $\nu$ fermionic zero modes weighted by the gauge field action
$S_\nu(A)$. The product is over all eigenvalues of the
Dirac operator. For fermions in the adjoint representation, the doubly
degenerate eigenvalues count only once (Majorana fermions)
\cite{LEUTWYLER-SMILGA-1993}.
All sum rules that can be obtained from chiral perturbation theory follow
by equating the expansion in powers of $m^2$ of the
QCD partition function (1) and the low-energy effective partition function.
For $N_f = 1$, the fermion determinant can be expanded as
\be
 \prod_{\lambda_n > 0} (\lambda_n^2 + m^2) =
\prod_{\lambda_n > 0}\lambda_n^2
\left (1 + {\sum_{p=1}} m^{2p}\sum_{n_1\ne\cdots\ne n_p}
\frac 1{\lambda_{n_1}^2\cdots \lambda_{n_p}^2} \right ).
\ee
Therefore, one  finds sum rules for the following quantities
\be
S_p \equiv \left <\sum_{n_1\ne_2\ne\cdots\ne n_p}
\frac 1{\lambda_{n_1}^2\cdots \lambda_{n_p}^2} \right >,
\ee
where the average, which  includes the fermion determinant in the chiral limit,
is with respect to the QCD partition function.
For $N_f $ flavors, it is clear that sum rules for the quantities
\be
\left <\sum_{{\rm all}\,\, \lambda_k \,\,{\rm different}}
\frac 1{\lambda_{\alpha_1}^{2N_f}\cdots\lambda_{\alpha_{n_{N_f}}}^{2N_f}
\lambda_{\beta_1}^{2N_f-2}\cdots\lambda_{\beta_{n_{N_f-1}}}^{2N_f-2}\cdots
\lambda_{\omega_1}^2\cdots \lambda_{\omega_{n_1}}^2}\right >.
\ee
follow from a similar expansion of the product
\be
 \prod_{\lambda_n > 0}  \prod_{f=1}^{N_f} (\lambda^2_n + m_f^2).
\ee
Notice that in the thermodynamic limit these sum rules are only sensitive
to the spectrum near zero virtuality. In other words, they reflect
on the microscopic correlations of the spectrum, $i.e.$, correlations between
eigenvalues near zero  on the scale $1/V_4$.

\vskip 1.5cm
\noindent{\bf 3. Random matrix theory}
\vskip 0.5 cm
In this paper we calculate the sum rules, not via the chiral perturbation
expansion of the effective field theory but rather from chiral random matrix
theory, which, invoking universality arguments
\cite{SHURYAK-VERBAARSCHOT-1993,VERBAARSCHOT-1994A},
describes the spectrum of the QCD Dirac operator near zero virtuality.
In this case the partition function is defined by
\be
Z_{\beta,\nu}(m) = \int {\cal D}T P_\beta(T)\prod_f^{N_f}\det \left (
\begin{array}{cc} m_f & iT\\
                 iT^\dagger & m_f
\end{array}\right )
\ee
where $T$ has the symmetries of the corresponding Dirac operator and the
masses are in the chiral limit ($m_f\rightarrow 0$). As discussed above,
depending on the universality class the matrix $T$
is real ($\beta = 1$, chGOE), complex ($\beta = 2$, chGUE)
or quaternion real
($\beta = 4$, chGSE). In the latter case
the square root of the fermion determinant appears in (6).
The matrix $T$ is a rectangular $n\times m$ matrix with $|n-m| = \nu$
(for definiteness we take $m > n$), so that the 'Dirac operator' in (6) has
exactly $\nu$ zero modes.
The function $P(T)$ is chosen gaussian:
\be
P_\beta(T) = \exp\left (
{-\frac{\Sigma^2\beta n}{2}\sum_{k=1}^n \lambda_k^2}\right).
\ee
In this normalization, the chiral condensate equals $\Sigma$ in each of the
three random matrix ensembles.
The total number of modes is $N\equiv 2n$. The latter quantity
is identified with the volume of space time.
The joint probability density of the eigenvalues
of the Dirac operator is obtained by transforming to integration variables in
which $T$ is diagonal.
For $ N_f$ flavors and topological charge $\nu$  it is given by
\be
\rho_\beta(\lambda_1, \cdots, \lambda_n) = C_{\beta, n}
\prod_{k,l} |\lambda_k^2 -\lambda_l^2|^\beta \prod_{k}
\lambda_k^{2N_f +\beta\nu+\beta -1}
\exp\left ({-\frac{n\beta\Sigma^2}{2} \sum_k \lambda^2_k}\right ),
\ee
where $C_{\beta,n}$ are normalization constants.

The expectation values (4) are calculated with respect
to the joint probability density (8). To evaluate the integrals,
it is convenient to introduce new integration variables
\be
\mu_k = \frac {\beta n\Sigma^2}{2} \lambda_k^2,
\ee
resulting in the joint spectral density (up to a constant)
\be
\rho_{\beta,a}(\mu_1, \cdots, \mu_n) d\mu_1 \cdots d\mu_n = \prod_{k<l}
|\mu_k-\mu_l|^\beta \prod_k \mu_k^a
\exp(-\sum_k \mu_k) d\mu_1\cdots d\mu_n,
\ee
where
\be
a = \frac {(2N_f -2 + \beta + \beta \nu)}2.
\ee
Note that for $N_f= 1$ we have the following remarkable property
\be
\rho_{\beta,a}(\beta\mu_1, \cdots, \beta\mu_n) \sim
\left [\rho_{\beta=1,a}(\mu_1, \cdots, \mu_n) \right ]^\beta.
\ee

\vskip 1.5cm
\noindent{\bf 4. Recursion relations for Selberg's integral}
\vskip 0.5 cm

The evaluation of the spectral sum rules can be reduced to the calculation
of moments
\be
\left < \mu_1^{p_1} \cdots \mu_n^{p_n} \right >_{\rho_{\beta,a}}.
\ee
Because all integration variables occur symmetrically,
they only depend on the partitioning of the powers and will be denoted by
\be
H_{\beta,a}(n_p, n_{p-1}, \cdots, n_0), \qquad (\sum_k n_k = n).
\ee
Here, $n_k$ is the number of times the power $k$ occurs in (13).
By integrating $\partial_{\mu_1}\rho_{\beta,a}(\mu_1, \cdots, \mu_n)$
we find the recursion relation
\be
H_{\beta,a}(n_p, n_{p-1}, \cdots, n_0) &=&
(p+a)H_{\beta,a}(n_p - 1, n_{p-1}+1, \cdots n_0)
\nonumber\\
&+& \beta \sum_k \left < \frac{\prod_{i=1}^{n_p}\mu_i^p
{\prod_{j=1}^{n_p}}\mu_j^{p-1}
\cdots}{\mu_1 - \mu_k}\right>_{\rho_{\beta,a}}
\ee
The last term can be evaluated by exploiting that the the joint probability
density is a symmetric function with respect to all variables. If we
denote $\int dx_1 dx_2 S(x_1,x_2) F(x_1,x_2)$ by $\langle F\rangle_S$,
we can derive for a symmetric function $S(x_1,x_2)$ and even values of $q$
the relations
\be
\left< \frac {x_1^p x_2^{p-q}}{x_1 - x_2}\right>_S = \sum_{l=1}^{q/2}
\left <x_1^{p-l} x_2^{p-q+l-1}\right >_S,
\ee
where we have used that $\langle x_1^k x_2^k/(x_1-x_2)\rangle_S = 0$. For odd
values of $q$ a similar relation can be obtained
\be
\left< \frac {x_1^p x_2^{p-q}}{x_1 - x_2}\right>_S = \frac 12
\left < x_1^{p-(q+1)/2} x_2^{p-(q+1)/2} \right >_S +
\sum_{l=1}^{(q-1)/2}
\left <x_1^{p-l} x_2^{p-q+l-1}\right >_S.
\ee
In this case the series terminates because
\be
\left <\frac{ x_1^{k+1} x_2^k}{x_1-x_2}\right>_S =
\frac 12\langle x_1^k x_2^k \rangle_S.
\ee
With the help of these relations all terms in the sum of (15) can be
expressed in the functions $H_{\beta,a}$.
As a result we obtain the recursion relation
\be
H_{\beta,a}(n_p, n_{p-1}, \cdots, n_0) &=& (p+a + \frac {\beta}2 n_{p-1}
+ \beta\sum_{q=2}^p n_{p-q})H_{\beta,a}(n_p - 1, n_{p-1}+1)\nonumber \\
&+& \frac {\beta}2\sum_{k=1}^{[(p-1)/2]} n_{p-(2k+1)}
H_{\beta,a}(n_p -1, n_{p-k-1}+2, n_{p-(2k+1)} -1)\nonumber\\
&+&\beta\sum_{q=4}^p n_{p-q}\sum_{l=2}^{[q/2]}
H_{\beta,a}(n_p -1, n_{p-l}+1,n_{p-q+l-1}+1,n_{p-q} -1),\nonumber\\
\ee
where the largest integer smaller than $x$ is denoted by $[x]$, and we used
the notation that all arguments of $H$ not shown explicitly are the same as
in the l.h.s. of the equation.
In general, this recursion relation is very complicated and an analytical
solution seems illusive. However for $p = 1, \,2$ a linear recursion
relation is obtained which can be worked out analytically. For $p = 1$ the
result is \cite{MEHTA-1991}
\be
H_{\beta,a}(n_1, n_0) &=& (a +1+ \frac {\beta}{2} n_0)
H_{\beta,a}(n_1 -1, n_0)\nonumber \\
            &=& \prod_{k=0}^{n_1-1}(a +1+ \frac {\beta}{2} (n_0 +k) )
H_{\beta,a}(0, n_0+n_1).
\ee
For $p = 2$, the result is only slightly more complicated \cite{MEHTA-1991}:
\be
H_{\beta,a}(n_2, n_1,n_0) &=& (a+2+\frac{\beta}2 n_1 +\beta n_0)
                      H_{\beta,a}(n_2-1, n_1+1,n_0)\nonumber\\
                &=& \prod_{k =0}^{n_2-1}(a+2+\frac{\beta}2 (n_1+k) +\beta n_0)
                      H_{\beta,a}(0, n_1+n_2,n_0)\nonumber\\
                &=& \prod_{k =0}^{n_2-1}(a+2+\frac{\beta}2 (n_1+k) +\beta n_0)
    \nonumber\\ &\times&
           \prod_{l =0}^{n_1+n_2-1}(a+1+\frac{\beta}2 (n_0+l))
                 H_{\beta,a}(0, 0,n_0+n_1+n_2).\nonumber\\
\ee
To obtain the last equality we have used the relation for $p = 1$.

\vskip 1.5cm
\noindent{\bf 5. Results for the spectral sum rules}
\vskip 0.5 cm
It is now straightforward to evaluate $all$ sum rules for one and two
flavors. First, we consider the case $N_f = 1$. Since all
eigenvalues occur symmetrically in (3) we have
\be
S_p = \left ( \begin{array}{c} n \\ p \end{array} \right )
\left <\frac 1{\lambda_1^2\cdots \lambda_{_p}^2} \right >_{\rho_\beta}
\ee
In terms of the integration variables (9), the sum rules become
\be
S_p = \left ( \begin{array}{c} n \\ p \end{array} \right )
\left( \frac{\beta n}{2\sigma^2} \right )^p
\frac{\left< \mu_1 \cdots \mu_{n-p}\right >_{\rho_{\beta,a-1 }}}
{\left< \mu_1 \cdots \mu_{n}\right >_{\rho_{\beta,a-1} }}.
\ee
In order to make contact with Selberg's integral, the new average is
with respect to $\rho_{\beta, a-1}$. Application of (20) yields
\be
S_p = \frac 1{\prod_{k=0}^{p-1} ({\beta +\beta\nu}+ 2(N_f-1) +\beta k)}
\left ( \begin{array}{c} n \\ p \end{array} \right )
\left( {\beta n\Sigma^2} \right )^p.
\ee
Remarkably, for $N_f = 1$ the variable $\beta$ drops out of the equation
and the sum rule simplifies to
\be
S_p = \frac {\nu !}{(p+\nu)!}
\left ( \begin{array}{c} n \\ p \end{array} \right )
(n\Sigma^2)^p.
\ee
This result constitutes the resolution of the paradox posed in the
introduction. In fact, we have constructed a one-parameter family of spectra
which all give rise to the same one-flavor sum rules.
Using that the total number of modes is $N=2n$
the large $n$ limit of this  sum rule is given by
\be
S_p \sim
\frac {\nu!(N\Sigma)^{2p}}{p!(p+\nu)!2^{2p}},
\ee
which
coincides with a result obtained
in \cite{SMILGA-1992,LEUTWYLER-SMILGA-1993}
on general grounds involving the anomalous and chiral structure of  QCD.
For an arbitrary number of flavors, the large $n$ limit of (24) is given by
\be
S_p = \frac{(N^2\Sigma^2)^p}{2^{2p} p!} \frac{\Gamma(\nu+1+2(N_f-1)/\beta)}
{\Gamma(\nu+p+1+2(N_f-1)/\beta)},
\ee
which again agrees with previous work \cite{LEUTWYLER-SMILGA-1993}.

The most general sum rule for $N_f =2 $ is given by
\be
S_{pq} = \left < \sum_{{\rm all}\,\, \lambda_i\,\, {\rm different}}
\frac 1{\lambda_{m_1}^4
\cdots \lambda_{m_p}^4\lambda_{n_1}^2\cdots\lambda_{n
_q}^2}\right>.
\ee
These sum rules are also well-defined for any number of flavors $N_f >2$,
and for less than two flavors only under certain conditions on $\nu$ and
$\beta$. Therefore, we will evaluate the sum (27) for an arbitrary
value of $N_f$, under the assumption that the integral is convergent.
To evaluate the integrals we change variables as in the case of $N_f = 1$.
Exploiting the symmetry of the integration variables we obtain
\be
S_{pq} = N_{pq}\frac {\left< \prod_{k=1}^{n-p-q} \mu_k^2
\prod_{l=n-p-q+1}^{n-p} \mu_k \right >_{\rho_{\beta,a-2}}}
{\left< \prod_{k=1}^{n} \mu_k^2 \right >_{\rho_{\beta,a-2}}}
\left ( \frac {\beta n \Sigma^2}{2}\right )^{q+2p},
\ee
where the total number of terms in (27) is denoted by
\be
N_{pq}=\frac {n!}{p! q! (n-p-q)!}
\ee
The integrals in (28) follow immediately from the recursion (21).
The result can be expressed
in terms of $\Gamma$-functions:
\be
S_{pq} = N_{pq}\frac {  \Gamma(\alpha + \frac 2{\beta}) \Gamma(\alpha+p)
                        \Gamma(\alpha + \frac 2{\beta} + p + n)}
                     {\Gamma(\alpha + \frac 2{\beta}+q+2p) \Gamma(\alpha)
                        \Gamma(\alpha + \frac 2{\beta} + n)}
                   \left (  n^2\Sigma^2 \right )^{q+2p},
\ee
where
\be
\alpha = \nu + 1 + \frac{2(N_f -2)}{\beta}.
\ee
We observe that for $p = 0$ these sum-rules reduce to the case $N_f = 1$.
For $p\ne 0$ and $N_f = 1$ the $\beta-$dependence does not drop form (30).
This is consistent with the fact that in this case the sum rules cannot be
obtained from a chiral expansion of the partition function.
For $N_f \ne 1$ the result depends on $\beta$.
The thermodynamic limit of the sum-rules can be obtained with the help
of Stirling's formula. This yields
\be
S_{pq} \sim \frac { \left ( N^2\Sigma^2 \right )^{q+2p}}
                  {4^{q+2p} p! q!}
 \frac{\Gamma(\alpha + \frac 2{\beta})\Gamma(\alpha)}
      {\Gamma(\alpha + \frac 2{\beta} + q + 2p)\Gamma(\alpha+p)}.
\ee
Two cases that have been considered before in
\cite{LEUTWYLER-SMILGA-1993}, namely $\beta =2$, $p = 1$, $q =0$ and
$\beta =2$, $p = 0$, $q =2$, are reproduced.

\vskip 1.5cm
\noindent{\bf 6. Discussion}
\vskip 0.5 cm
In conclusion, we have evaluated all sum rules that
follow from a chiral expansion of the partition function
for one and two flavors. Each sum rule has been calculated for
an arbitrary number of flavors and in a sector with a given topological charge.
For $one$ flavor we have found that all sum rules that can be derived
from the effective low-energy partition function coincide for each of the
three universality classes. This in spite of the fact that
the spectral correlations are different. This resolves the paradox posed
in the introduction of this letter. For two or more flavors, both
the effective theory and the Leutwyler-Smilga
sum rules are different for different values of $\beta$. However, we expect
that also in this case the spectrum is not determined uniquely determined
by the effective low-energy theory.

To see the connection between the spectral density near zero virtuality
and the low-energy effective theory, consider the partition function
\be
Z_\nu^k(m,z) =  \left<\prod_{\lambda_n > 0} (\lambda_n^2 + z^2)^k
\prod_{f=1}^{N_f} \prod_{\lambda_n > 0} (\lambda_n^2 + m_f^2)m_f^\nu\right
>_{S_\nu(A)}.
\ee
The introduction of $k$ replicated flavors allows us to extract information
on the spectral density.
For $\nu = 0$, for example, we have the relation
\be
2\pi\rho(iz) = \lim_{k\rightarrow 0} \frac 1k \frac d{dz}
Z^k_{\beta,\nu=0}(m,z)
\ee
To perform the $k\rightarrow 0$ limit of the effective partition function, we
first evaluate it for any integer value of $k$ and take the limit after
analytical continuation. This implies that in order to obtain the full
spectral density for $N_f$ flavors, we do not only need the effective
theory for $N_f$ flavors, but for any larger number of flavors (with a
different mass) as well.
Since the effective partition function for an arbitrary number of flavors
with equal masses is known \cite{LEUTWYLER-SMILGA-1993}, this program
may be feasible.

On the other hand, using techniques described in
\cite{VERBAARSCHOT-1994A,SHURYAK-VERBAARSCHOT-1993}
\cite{VERBAARSCHOT-WEIDENMUELLER-ZIRNBAUER-1985}, the spectral density
near zero virtuality may be related to an effective theory based
on graded cosets. At the moment it is not clear which approach will be most
successful.

\vglue 0.6cm
{\bf \noindent  Acknowledgements \hfil}
\vglue 0.4cm
The reported work was partially supported by the US DOE grant
DE-FG-88ER40388. I would like to thank A. Smilga for pointing out the
special status of $N_f = 1$.

\vfill
\eject
\newpage
\setlength{\baselineskip}{15pt}

\bibliographystyle{aip}

\end{document}